\documentclass[letterpaper]{article} 
\usepackage{aaai2026}  
\usepackage{times}  
\usepackage{helvet}  
\usepackage{courier}  
\usepackage[hyphens]{url}  
\usepackage{graphicx} 
\usepackage{natbib}  
\usepackage{caption} 
\usepackage{algorithm}
\usepackage{algorithmic}

\newcounter{checksubsection}
\newcounter{checkitem}[checksubsection]
\usepackage{amsmath}
\usepackage[table]{xcolor}
\usepackage{colortbl}
\usepackage[capitalize]{cleveref} 
\usepackage{multirow}
\usepackage{booktabs}

\usepackage{newfloat}
\usepackage{listings}
\DeclareCaptionStyle{ruled}{labelfont=normalfont,labelsep=colon,strut=off} 
\floatstyle{ruled}
\newfloat{listing}{tb}{lst}{}
\floatname{listing}{Listing}
\title{COSMOS: Coherent Supergaussian Modeling 
with Spatial Priors for Sparse-View 3D Splatting}
\author{
    Chaeyoung Jeong\textsuperscript{\rm 1}, Kwangsu Kim\textsuperscript{\rm 2}\thanks{Corresponding author.} 
}
\affiliations{
    \textsuperscript{\rm 1}Department of Artificial Intelligence, Sungkyunkwan University\\

    \textsuperscript{\rm 2}Department of Computer Science and Engineering, Sungkyunkwan University\\
    pcc1016@skku.edu, kim.kwangsu@skku.edu
}

\usepackage{bibentry}

\begin{document}

\maketitle

\begin{abstract}
3D Gaussian Splatting (3DGS) has recently emerged as a promising approach for 3D reconstruction, providing explicit, point-based representations and enabling high-quality real-time rendering.
However, when trained with sparse input views, 3DGS suffers from overfitting and structural degradation, leading to poor generalization on novel views. This limitation arises from its optimization relying solely on photometric loss without incorporating any 3D structure priors. To address this issue, we propose Coherent supergaussian Modeling with Spatial Priors (COSMOS).
Inspired by the concept of superpoints from 3D segmentation, COSMOS introduces 3D structure priors by newly defining supergaussian groupings of Gaussians based on local geometric cues and appearance features.
To this end, COSMOS applies inter-group global self-attention across supergaussian groups and sparse local attention among individual Gaussians, enabling the integration of global and local spatial information. These structure-aware features are then used for predicting Gaussian attributes, facilitating more consistent 3D reconstructions. Furthermore, by leveraging supergaussian-based grouping, COSMOS enforces an intra-group positional regularization to maintain structural coherence and suppress floaters, thereby enhancing training stability under sparse-view conditions.
Our experiments on Blender and DTU show that COSMOS surpasses state‑of‑the‑art methods in sparse‑view settings without any external depth supervision.
\end{abstract}


\section{Introduction}

\begin{figure}[t]
\centering
\includegraphics[width=1\columnwidth]{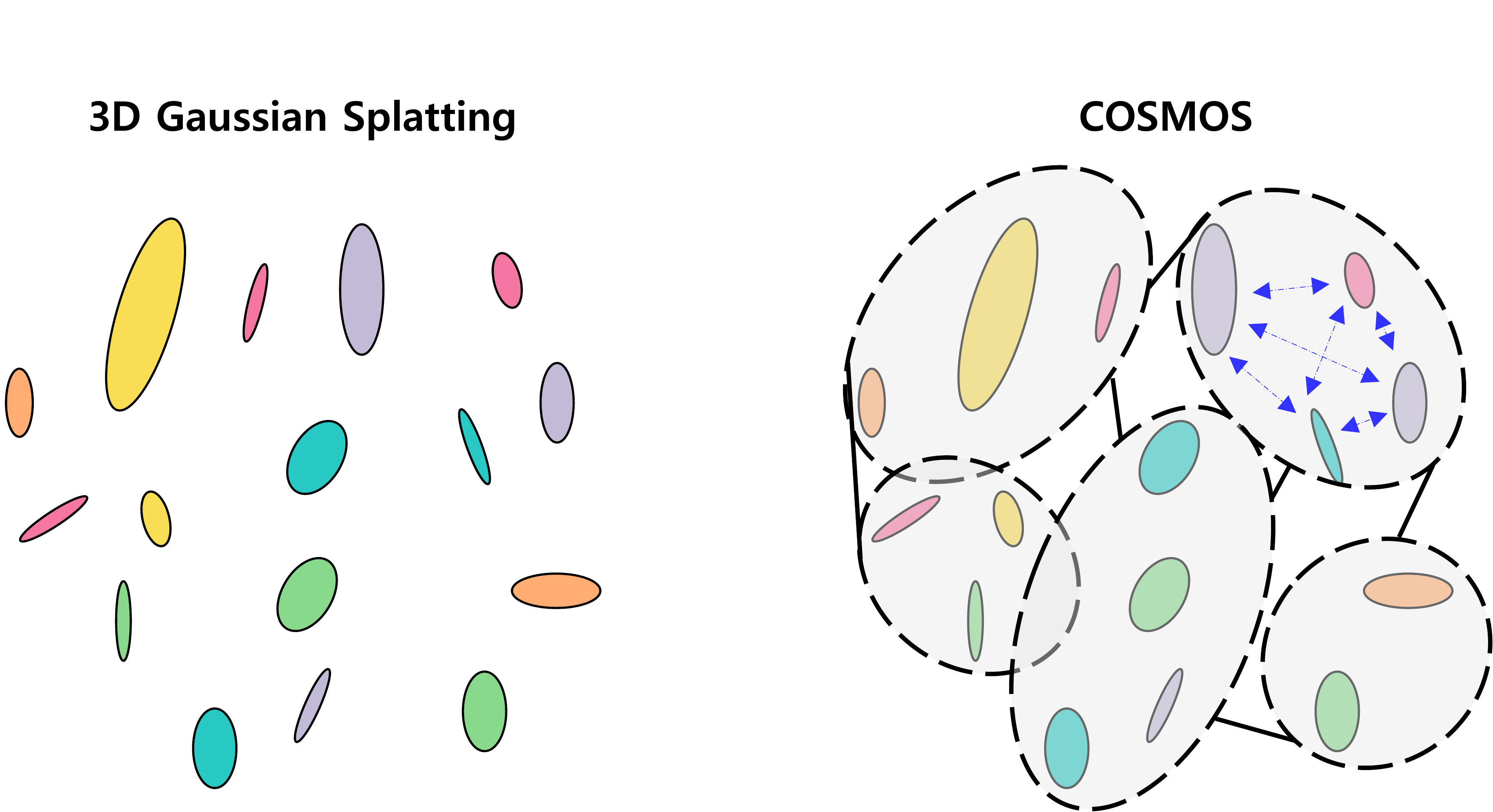} 
\caption{\textbf{Comparison of the Optimization Process between 3DGS and COSMOS.} While 3DGS learns each Gaussian independently to fit the training views, COSMOS leverages supergaussian grouping to inject spatial priors during optimization. Through inter‑group learning, it captures the global 3D structural context, while intra‑group regularization prevents Gaussians within similar geometric structures from diverging in different directions.}
\label{fig1}
\end{figure}

3D scene reconstruction aims to learn a 3D representation capable of synthesizing novel views from 2D images. This technology can reduce reliance on manual 3D modeling and has broad applicability across industries such as film production, robotic navigation, cultural heritage preservation, VR/AR and computer graphics.
Recently, learning-based 3D reconstruction methods—such as Neural Radiance Fields (NeRF) \cite{mildenhall2021nerf}—have garnered considerable attention for their capacity to model continuous 3D scenes. Among these methods, 3D Gaussian Splatting (3DGS) \cite{kerbl20233d} stands out because of its point-based, explicit representation, which is compatible with conventional 3D rendering pipelines and capable of high-quality novel view synthesis.

Despite its advantages in dense-view settings, optimizing 3DGS with sparse input views remains challenging. In practice, acquiring dense multi-view images for a single scene is often prohibitively costly and time-consuming, highlighting the need for models that remain robust under limited-view conditions. The primary reason for 3DGS's limited generalization under sparse inputs is its reliance on simple pixel-wise photometric loss between rendered and training images without leveraging explicit 3D structural priors. As a result, 3DGS tends to overfit and fails to generalize to unseen viewpoints.

To address these limitations, recent studies have introduced prior knowledge to improve novel view synthesis under sparse input conditions \cite{mihajlovic2024splatfields,xiong2023sparsegs}. A representative approach involves using pseudo-depth maps generated by auxiliary or pretrained models to regularize the geometry \cite{li2024dngaussian,deng2022depth,xu2024mvpgs,chung2024depth,zhu2024fsgs}. While these approaches aim to preserve structural consistency under sparse input conditions, their performance is highly dependent on the accuracy of the predicted depths, and the absolute scales of depth values vary across views, causing misalignment \cite{park2025dropgaussian, xu2025dropoutgs}.
To overcome these limitations, we propose an alternative 3D prior that provides structural cues without relying on depth supervision.

We propose \textbf{CO}herent \textbf{S}uperGaussian \textbf{MO}deling with \textbf{S}patial Priors (COSMOS), a framework that preserves the explicit, point-based representation of 3DGS while overcoming its limitations. Inspired by superpoints \cite{landrieu2018large} in semantic segmentation, COSMOS redefines the grouping criteria to suit the 3DGS context. Specifically, it groups Gaussians into supergaussians based on local geometric relations and appearance features, thereby injecting structured 3D priors into the learning process.

These supergaussians serve two primary purposes. First, compressing thousands of Gaussians into only a few tens groups, COSMOS enables efficient global self-attention \cite{vaswani2017attention} across supergaussian groups, as well as distance-based sparse self-attention \cite{child2019generating} among individual Gaussians. This design allows the network to learn 3D feature vectors that capture both global and local spatial structure. These features guide Gaussian-attribute prediction and promote coherent 3D geometry under sparse input conditions. Second, intra-group regularization suppresses divergence among Gaussians belonging to the same structural region, thereby mitigating the occurrence of floaters. By injecting these supergaussian-based 3D priors into the model (Figure~\ref{fig1}), COSMOS achieves robust and structurally coherent novel view synthesis. It also overcomes the sparse-view limitations of 3DGS.

In summary, our contributions are as follows:
\begin{enumerate}
    \item We propose COSMOS, an attribute prediction framework that introduces a novel 3D structural prior through our newly defined inter‑group supergaussian representation. COSMOS integrates global supergaussian embeddings with local Gaussian features to achieve structurally consistent predictions of Gaussian attributes.
    \item Without relying on external depth supervision, COSMOS introduces an intra-group positional regularization to suppress floaters and enhance structural consistency under sparse-view conditions.
    \item COSMOS achieves superior novel view synthesis performance under extremely sparse input conditions.
Compared to recent state-of-the-art (SOTA) methods in 3D scene reconstruction, COSMOS demonstrates strong performance on the Blender and DTU datasets using as few as three input views. Experimental results show that our approach effectively reconstructs complex structures and high-frequency details even with minimal supervision.
\end{enumerate}

\section{Related Work}

\subsection{Neural Representations for 3D reconstruction}
3D reconstruction aims to synthesize images from unobserved camera viewpoints using a set of observed views.
Traditional approaches such as Structure-from-Motion (SfM) \cite{schonberger2016structure} and Multi-View Stereo (MVS) have long been studied but suffer from limitations including complex pipelines and fragile matching procedures. To overcome these challenges, recent research has increasingly focused on learning-based methods for novel view synthesis.

Among these, NeRF \cite{mildenhall2021nerf} models a scene as an implicit continuous function that maps 3D coordinates and viewing directions to color and density, producing photorealistic novel views.
However, its implicit formulation lacks explicit geometry for direct manipulation and requires expensive per‑ray sampling and integration, making training slow \cite{li2023nerfacc, jambon2023nerfshop}.


In contrast, the recently introduced 3DGS \cite{kerbl20233d} offers an explicit, point-based 3D representation that enables fast training and compatibility with traditional 3D rasterization pipelines.
Each 3D Gaussian encodes position, covariance, color, and opacity, and is rendered as an ellipsoidal volume that can overlap with multiple pixels, leading to efficient and photorealistic rendering.
Thanks to its simplicity and efficiency, 3DGS has rapidly become a central foundation for follow‑up research, inspiring numerous extensions for memory efficient, dynamic scenes, and high‑fidelity rendering \cite{hamdi2024ges,lee2024compact,sun20243dgstream,wu20244d,yu2024mip}.

Nevertheless, 3DGS still requires a large number of input views to achieve reliable novel view synthesis. When trained under sparse input conditions, it tends to fail to preserve object integrity and exhibits significant performance degradation in unseen views.

\subsection{Few-shot 3D reconstruction}
Few-shot 3D reconstruction, which aims to perform novel view synthesis from a limited set of input views, has been actively studied for years \cite{yu2021pixelnerf,niemeyer2022regnerf,xu2022sinnerf,xu2025dropoutgs,somraj2023simplenerf,yang2024gaussianobject,johari2022geonerf}.
DietNeRF \cite{jain2021putting} improves performance by injecting semantic information into the reconstruction process using CLIP \cite{radford2021learning}, a large-scale image-text alignment model.
FreeNeRF \cite{yang2023freenerf} identifies the high-frequency components of NeRF’s positional encoding as a key contributor to overfitting. It improves generalization by partially masking these positional encodings during training.
Other works have enhanced novel view synthesis by guiding the model with depth-based regularization to encourage more plausible geometry. This approach constrains the scene’s appearance using predicted depths \cite{li2024dngaussian,deng2022depth,xu2024mvpgs}.
However, such approaches rely on auxiliary or pretrained models to obtain depth supervision, limiting both flexibility and scalability.

In contrast, COSMOS instead regularizes the appearance without any depth supervision. This strategy mitigates overfitting and the occurrence of floaters, ultimately improving novel view synthesis under sparse input settings.


\subsection{Structured Group Representations}
The concept of superpoints originates from the idea of superpixels \cite{achanta2012slic} in 2D image processing, where geometrically homogeneous point sets are grouped into a single unit. This approach was first introduced in the context of 3D semantic segmentation. Superpointgraph \cite{landrieu2018large} estimates the geometric properties of each superpoint by computing each superpoint’s coordinate covariance and analyzing its eigenvalues to derive descriptors (length, planarity, volumetricity, etc.) Based on these descriptors, the spatial relationships between superpoints are modeled as a graph, with edges encoding various geometric attributes such as relative position offsets, volume differences, and shape ratios. This representation allows for efficient processing of large-scale point clouds while enhancing structural expressiveness.
Superpoints have since been widely adopted across various large-scale point cloud tasks, leading to a surge in related research \cite{robert2023efficient,kolodiazhnyi2024oneformer3d,hui2021superpoint}.
COSMOS builds upon these prior works by incorporating the superpoint concept into few-shot 3DGS, introducing so-called supergaussians in 3DGS. Gaussians are grouped into supergaussians according to our defined grouping criterion based on local geometric and appearance features, injecting spatial bias and structural priors into the 3DGS framework. By leveraging these 3D priors, COSMOS achieves more robust generalization in few-shot 3D reconstruction scenarios.

\section{Methodology}

\begin{figure*}[t]
  \centering
  \includegraphics[width=\linewidth]{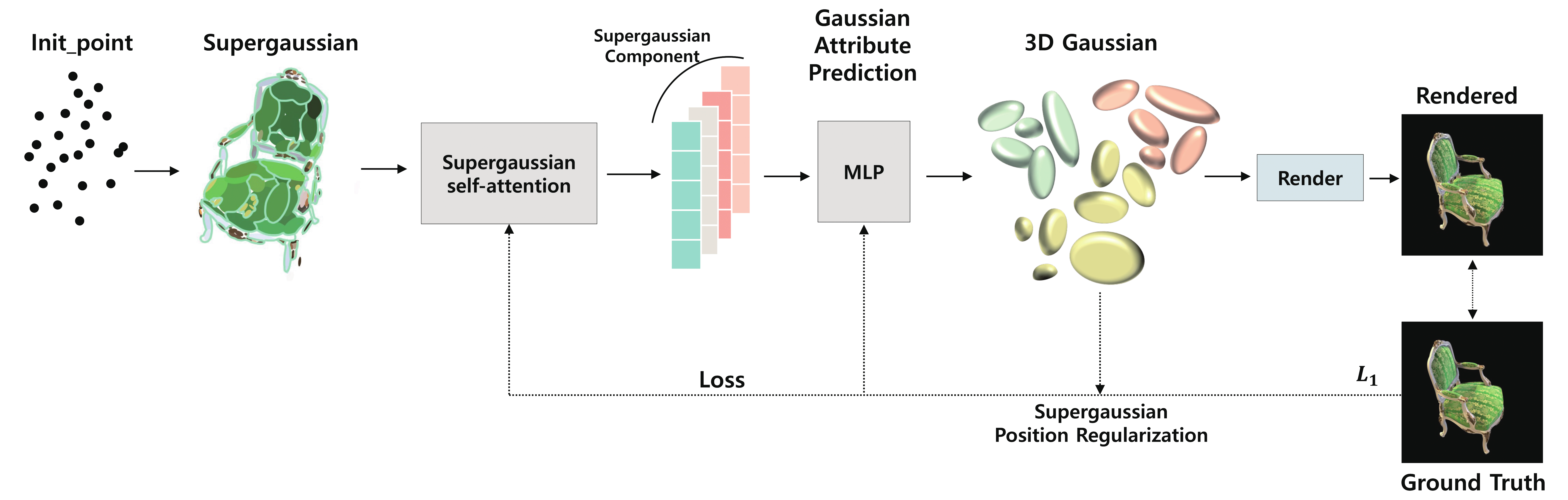}
  \caption{\textbf{Architecture of COSMOS.} To incorporate 3D priors, we initially group Gaussians into supergaussians based on geometric cues. A self-attention module operating at the supergaussian level is then employed to generate 3D feature vectors, enabling each Gaussian to be trained in a structurally informed manner rather than independently. Furthermore, we introduce a regularization term that constrains the positional deviation of Gaussians within the same supergaussian, which helps suppress floaters and alleviates structural collapse under sparse input conditions.}
  \label{fig2}
\end{figure*}

Our approach is built upon 3DGS, which enables explicit, point-based 3D reconstruction (\cref{sec:3dgs_pre}).   COSMOS begins by grouping 3D Gaussians into supergaussians based on superpoint-inspired grouping (\cref{sec:supergaussian}).
We then use a transformer architecture to compute global embeddings for each supergaussian, and local embeddings for individual Gaussians which are fed into the Gaussian attribute prediction module (\cref{sec:sgsa}).
Finally, the model is optimized using a supergaussian-level regularization loss (\cref{sec:sgpr}).
Figure~\ref{fig2} illustrates the overall architecture of COSMOS. 

\subsection{Preliminaries for 3DGS}
\label{sec:3dgs_pre}
3D Gaussians are point-based primitives used to represent a 3D scene, where each Gaussian is parameterized by its position, covariance, color, and opacity. A Gaussian can be defined by a center position \( x \) and a 3D covariance matrix \( \Sigma \) in 3D space:

\begin{equation}
\scalebox{0.9}{$
G(x) = e^{-\frac{1}{2}(x)^T \Sigma^{-1}(x)}
$}
\end{equation}

Since the covariance matrix must be positive definite, direct optimization via gradient descent can easily result in invalid matrices. To ensure valid and stable optimization, each Gaussian's covariance is reparameterized as a combination of a rotation matrix and a scaling matrix. Specifically, a 3D scale vector \( S \) encodes the axis-aligned scaling and a unit-norm quaternion encodes the rotation \( R \).

\begin{equation}
\scalebox{0.9}{$
\Sigma = R S S^{T} R^{T}
$}
\end{equation}

After optimization, we project the Gaussians into camera space using the viewing transformation \( W \). Then we apply \( J \), the Jacobian of the projective transformation’s affine approximation, to the Gaussian parameters to enable rendering.

\begin{equation}
\scalebox{0.9}{$
\Sigma' = J W \Sigma W^{T} J^{T}
$}
\end{equation}

During rendering, each Gaussian is rasterized and contributes to multiple pixels along the ray. The final pixel color  \( C \) is computed by aggregating the projected attributes of \( N \) Gaussians along each ray:

\begin{equation}
C = \sum_{i \in N} \left( \alpha_i c_i \prod_{j=1}^{i-1} (1 - \alpha_j) \right)
\end{equation}

where \( c_i \) denotes the color of each point, and \( \alpha_i \) is computed by evaluating the 2D Gaussian obtained from projecting the 3D Gaussian with covariance $\Sigma$ onto the image plane, and scaling it by a learned per‑Gaussian opacity parameter.
3DGS supervises the model with an \( \mathcal{L}_1\) loss (and an \( \mathcal{L}_\text{ssim}\)) between the rendered image and the ground-truth image. The Gaussian attributes are optimized to minimize this discrepancy, improving scene fidelity.

\subsection{Supergaussian Grouping}
\label{sec:supergaussian}
To alleviate floaters and structural collapse in sparse-view settings, we begin by grouping Gaussians to introduce 3D priors into 3DGS. Specifically, we compute local geometric descriptors for each Gaussian, including linearity (\( \ell_i \)), scattering (\( s_i \)), verticality (\( v_i \)), and planarity (\( p_i \)) \cite{landrieu2018large}. 
These descriptors are concatenated with the Gaussian’s position (\( x_i \)), color (\( c_i \)), and scale (\( \sigma_i \)) to form the feature vector used for grouping. 
We then apply the \( \ell_0 \)-Cut pursuit algorithm \cite{landrieu2017cut} to partition the Gaussians into groups. This procedure requires only a few graph-cut iterations and efficiently produces supergaussians.

\begin{equation}
    f_i = \big[ \ x_i,\ c_i,\ \sigma_i,\ \ell_i,\ s_i,\ v_i,\ p_i]
    \label{eq:feature_concat}
\end{equation}

Here, \( i \) denotes the index of a Gaussian.
Unlike original superpoints, COSMOS additionally incorporates Gaussian scale as a grouping feature.
The scale of a Gaussian plays a critical role in the reconstruction process: smaller Gaussians tend to capture fine-grained details and high-frequency regions, whereas larger Gaussians contribute to the reconstruction of coarser global structures.
By including scale in the feature representation, the model can distinguish whether a given supergaussian group is intended to recover local detail or overall shape.

\begin{equation}
    S_g = \{\, i \mid k_i = g \,\}, \qquad g = 0,\dots,G-1
    \label{eq:group_assignment}
\end{equation}

$S_g$ denotes the set of Gaussian indices $i$ belonging to supergaussian $g$, 
where $G$ is the total number of supergaussian groups and $k_i$ indicates the group assignment of Gaussian $i$.
Following \cite{landrieu2018large}, we also compute each group’s spatial covariance and derive group-level linearity, planarity, verticality, and scattering descriptors to characterize its geometry.


\begin{figure*}[t]
  \centering

    \centering
    \includegraphics[width=\linewidth]{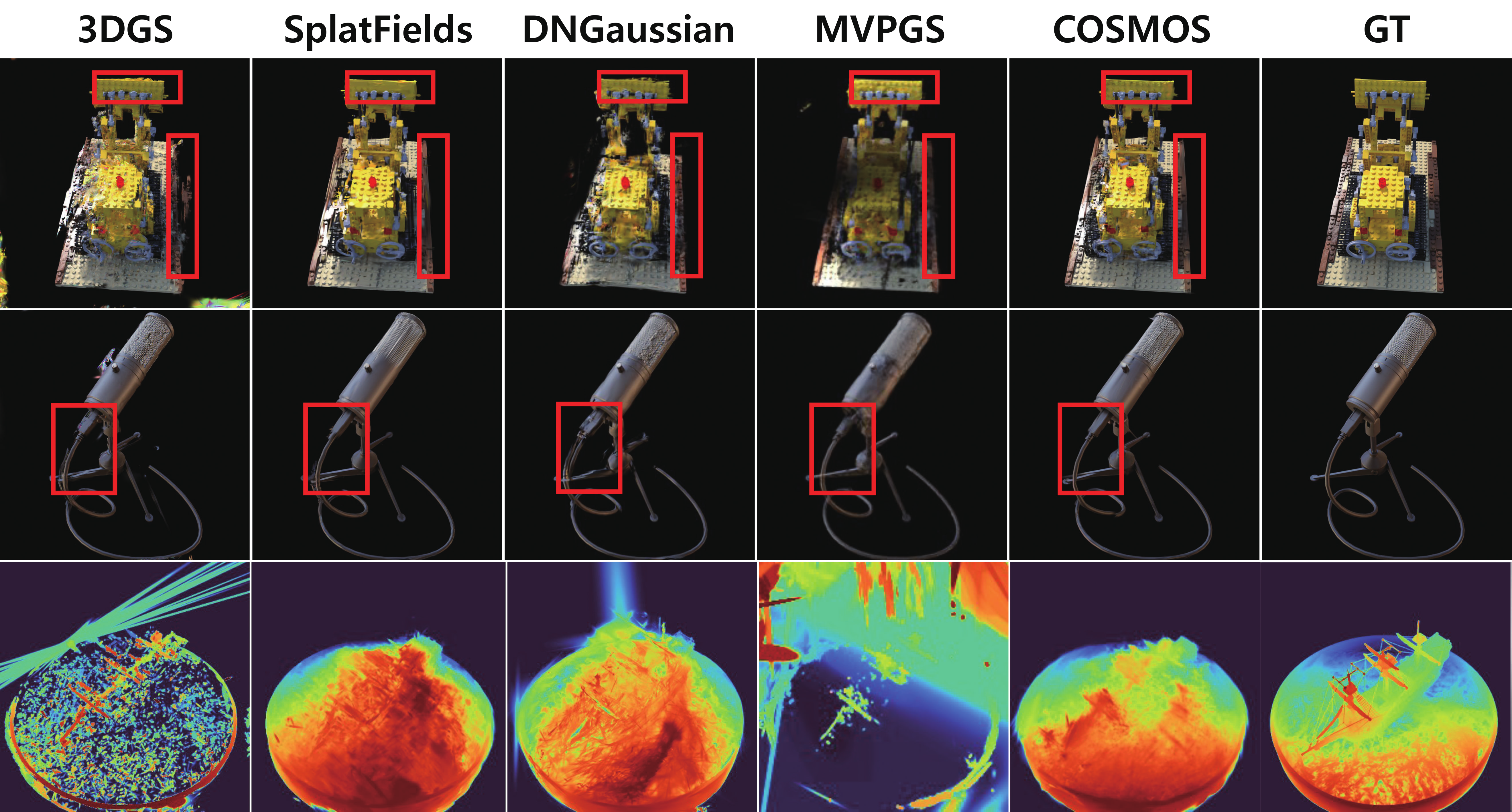}
    \caption{\textbf{Qualitative Comparison on Blender.} We train COSMOS and recent competitive models with 3 input views and render novel views for evaluation. COSMOS effectively suppresses floaters and restores fine high‑frequency details (see red boxes). The third‑row depth maps show that COSMOS reconstructs the smoothest and most continuous geometry without any depth supervision.}
    \label{fig:blender_visual}

\end{figure*}

\subsection{Supergaussian Self-Attention}
\label{sec:sgsa}
To inject spatial bias into the representation, self-attention can be effectively leveraged for point cloud embedding \cite{guo2021pct,yu2022point}. However, since a typical 3DGS scene is composed of tens to hundreds of thousands of Gaussians, directly applying global attention across all Gaussians is computationally prohibitive.
In contrast, COSMOS addresses this limitation by grouping large numbers of Gaussians into a few tens supergaussians, thus enabling efficient global self-attention at the group level.

We first perform group-wise max pooling to extract a representative feature \(s_g\) for each supergaussian, using the previously computed group assignments and per-Gaussian features.
 
 \begin{equation}
\begin{split}
 s_g\ = \max_{i\in S_g}f_i \quad \text{and} \quad s_g = [\,x_g;\;r_g\,] 
\end{split}
\end{equation}

 Positional encoding \( \gamma \) \cite{mildenhall2021nerf}  is then applied to the group center position \( x_g \) which we separate from the remaining features \( r_g \). After applying positional encoding to \( x_g \), it is concatenated with \( r_g \), and the combined features are passed through a linear projection layer \( W_\text{in}\), \( b_\text{in}\) to adjust their dimensionality.

\begin{equation}
\begin{split}
u_g &= 
W_{\text{in}}\;\text{Norm}\!\big([\gamma(x_g);r_g]\big)
+b_{\text{in}}
\end{split}
\end{equation}

The features are then normalized and fed through linear projection layers to obtain the query, key, and value embeddings for self-attention.

\begin{equation}
\begin{split}
q_g^h &= W^Q_h u_g, \quad
k_g^h= W^K_h u_g, \quad
v_g^h = W^V_h u_g
\end{split}
\end{equation}

Here, $W^Q_h$, $W^K_h$, and $W^V_h \in \mathbf{R}^{d_\text{model} \times d_h}$ are learned projection matrices that map the input features to queries \(q_g^h\), keys \(k_g^h\), and values \(v_g^h\), respectively. Using these queries, keys, and values, we compute scaled dot‑product attention across all supergaussians. The resulting attention weights capture the relevance between supergaussians, and the final embedding encodes the inter-group spatial context for each group.
The learned inter-group spatial features are then passed through an activation function and a final output projection layer, yielding a set of supergaussian embeddings that capture global structural context.

Aiming to capture high-frequency details and fine-grained spatial variations that cannot be fully learned through group-based dense attention, we introduce a sparse local attention mechanism \cite{child2019generating} in addition to global group-level attention.
In this design, each Gaussian is represented by a feature vector encoding both appearance and geometric attributes, and its 3D spatial coordinates are positionally encoded. Each Gaussian then attends to its ten nearest neighbors using a sparse transformer. Finally, the outputs from the global group attention and local sparse attention are concatenated to form a unified 3D feature vector for each Gaussian that integrates both global and local information.

To predict the final Gaussian attributes—namely position, orientation, scale, color, and opacity—COSMOS employs separate Residual MLPs \cite{mihajlovic2023resfields}, each dedicated to one specific attribute. These MLPs take as input the concatenation of the Gaussian’s spatial coordinates and the corresponding unified 3D feature, enabling structurally aware attribute prediction.

\begin{table*}[t]
    \centering
    \caption{\textbf{Quantitative Comparison on Blender.} 
    We mark the \textbf{best} and \underline{second best} results.}
    \label{tab:blender_quant}
    \begin{tabular}{lcccccc}
      \hline
      \multirow{2}{*}{Method} & \multicolumn{3}{c}{$N=3$} & \multicolumn{3}{c}{$N=6$} \\
      \cline{2-7}
       & \textbf{PSNR ↑} & \textbf{SSIM ↑} & \textbf{LPIPS ↓} 
       & \textbf{PSNR ↑} & \textbf{SSIM ↑} & \textbf{LPIPS ↓} \\
      \hline
      DietNeRF          & 10.47 & 0.57 & 0.42 & 14.10 & 0.66 & 0.34 \\
      FreeNeRF          & 16.75 & 0.72 & 0.23 & \textbf{23.67} & \textbf{0.87} & \textbf{0.11} \\
      MVPGS             & 16.64 & 0.66 & 0.30 & 21.07 & 0.78 & 0.17 \\
      DNGaussian        & 16.28 & 0.74 & 0.21 & 21.79 & 0.85 & \textbf{0.11} \\
      SplatFields       & \underline{18.46} & \underline{0.77} & \underline{0.20} & 22.26 & \underline{0.86} & \underline{0.15} \\
      3DGS              & 14.77 & 0.58 & 0.32 & 21.05 & 0.80 & 0.16 \\
      COSMOS (Ours)     & \textbf{18.96} & \textbf{0.79} & \textbf{0.19} & \underline{22.78} & 0.85 & \underline{0.15} \\
      \hline
    \end{tabular}
\end{table*}

 \begin{figure*}[t]
  \centering

    \centering
    \includegraphics[width=\linewidth]{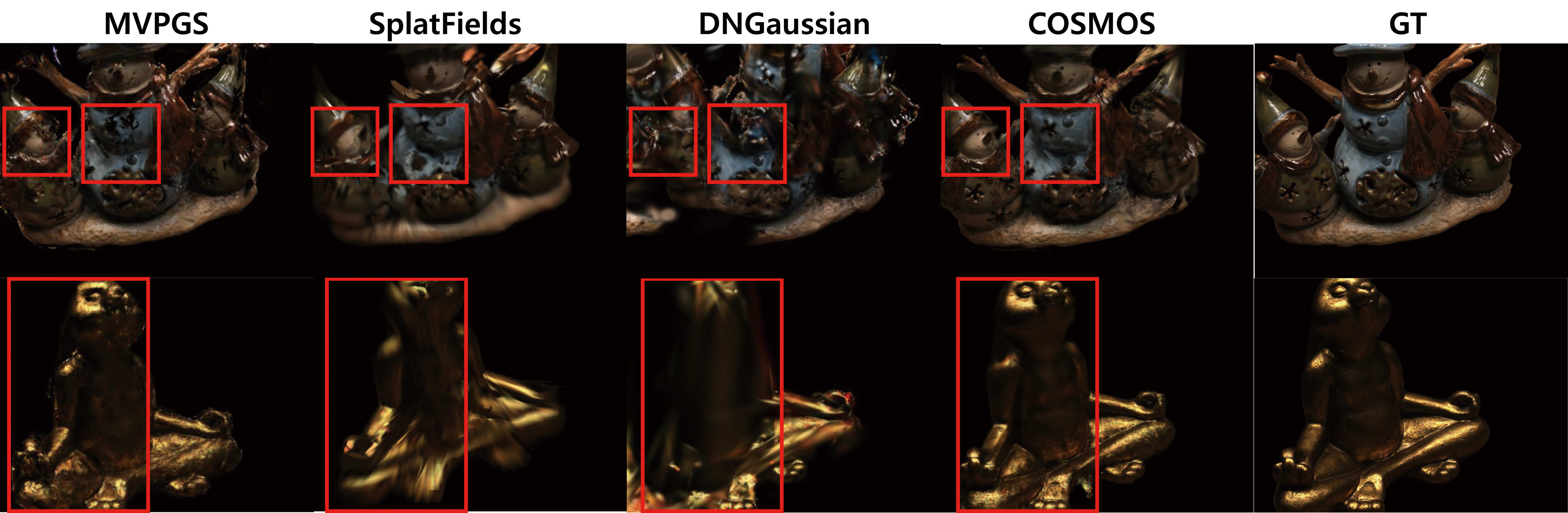}
    \caption{\textbf{Qualitative Comparison on DTU.} We train MVPGS, SplatFields, DNGaussian, and COSMOS using 3 input views and render novel views for evaluation. COSMOS demonstrates superior novel view synthesis performance, even on real-world datasets with complex textures and structures.}
    \label{fig:dtu_qual}

\end{figure*}

\subsection{Supergaussian Position Regularization}
\label{sec:sgpr}
To mitigate the tendency of Gaussians to overfit to training views, COSMOS introduces a supergaussian regularization that enforces intra-group attribute consistency. This regularization encourages Gaussians within the same group to learn coherent representations, improving generalization to unseen views.

Supergaussian position regularization consists of two regularization terms. First, it constrains the learning of Gaussian positions to prevent them from drifting too far from their neighboring Gaussians.

\newcommand{\R}{\mathbb{R}}
\begin{equation}
D_\text{avg}
= \frac{1}{N}\sum_{i=1}^{N}
\frac{1}{|\mathcal{N}(i)|+\varepsilon}
\sum_{j\in \mathcal{N}(i)}
\left\lVert x_i - x_j \right\rVert_2
\end{equation}

$\mathcal{N}(i)=\{\, j \mid (i,j)\in \mathcal{E} \,\}$ denotes the set of neighboring Gaussians of the \(i\)-th Gaussian, where $\mathcal{E}$ is the set of edges connecting neighboring Gaussians, and \( N \) is the total number of Gaussians.

The second term encourages each Gaussian to remain close to the center of its assigned supergaussian, preventing Gaussians within similar geometric structures from drifting in different directions. This promotes continuity and structural consistency during learning.
  
\begin{equation}
D_\text{ctr}
= \frac{1}{N}\sum_{i=1}^{N}
\left\|
x_i -
\frac{1}{|S_{k_i}|+\varepsilon}
\sum_{j \in S_{k_i}} x_j
\right\|_2^2
\end{equation}

$|S_{k_i}|$ represents the size of the supergaussian to which the \(i\)-th Gaussian  belongs. The overall position regularization loss is formulated as follows:

\begin{equation}
\mathcal{L}_{\text{pos}} = \lambda_1\,D_\text{avg} + \lambda_2\,D_\text{ctr}
\end{equation}

These constraints mitigate overfitting under sparse-view training, preventing Gaussians from drifting to spurious positions (which cause ‘floaters’ artifacts).

The final loss combines the image reconstruction term \( \mathcal{L}_1\) (with \( \mathcal{L}_\text{ssim}\)), the regularization term \( \mathcal{L}_\text{pos}\) and an optional mask-based loss \( \mathcal{L}_\text{mask}\) :

\begin{equation}
\mathcal{L}_{\text{total}} = \lambda \mathcal{L}_1 + (1 - \lambda) \mathcal{L}_{\text{SSIM}} + \lambda_{\text{pos}} \mathcal{L}_{\text{pos}} + \lambda_{\text{mask}} \mathcal{L}_{\text{mask}}
\label{eq:total_loss}
\end{equation}

Here, $\lambda$ balances \( \mathcal{L}_1\) and \( \mathcal{L}_\text{ssim}\). $\lambda_{pos}$ and $\lambda_{mask}$ weight the regularization terms.

\section{Experiments}

\subsection{Setup}
\paragraph{Datasets and Metrics}
We conduct experiments on the Blender dataset \cite{mildenhall2021nerf} provided by NeRF, and the real-world multi-view dataset DTU \cite{jensen2014large}. For Blender, we use two settings: using 3 or 6 training views, following the same view sampling protocol as previous works \cite{mihajlovic2024splatfields}. Evaluation is then performed on 200 test views, as in the original NeRF setting. For DTU, we adopt the same experimental setup as prior studies \cite{huang20242d,niemeyer2022regnerf}, using 3 training views for each scene.
We report PSNR, SSIM \cite{wang2004image}, and LPIPS \cite{zhang2018unreasonable} metrics, following standard evaluation protocols used in prior novel view synthesis studies.


\begin{table}[t]
    \centering
    \caption{\textbf{Quantitative Comparison on DTU ($N=3$).} 
    We mark the \textbf{best} and \underline{second best} results.}
    \label{tab:dtu_quant}
    \begin{tabular}{lccc}
        \toprule
        Method & PSNR ↑ & SSIM ↑ & LPIPS ↓ \\
        \midrule
        FreeNeRF        & 10.10 & 0.24 & 0.63 \\
        MVPGS           & 20.71 & 0.75 & \underline{0.19} \\
        DNGaussian      & 16.07 & \textbf{0.80} & \textbf{0.15} \\
        SplatFields     & \underline{21.07} & \underline{0.78} & 0.20 \\
        3DGS            & 19.40 & 0.54 & 0.48 \\
        COSMOS (Ours)   & \textbf{21.86} & 0.76 & 0.23 \\
        \bottomrule
    \end{tabular}
\end{table}

\paragraph{Implementation details}

All experiments are performed on a single NVIDIA A6000 GPU. We build our implementation on SplatFields \cite{mihajlovic2024splatfields}, removing its encoder component. We train each model for 12,000 iterations. Grouping into supergaussians is performed after the first 100 iterations of training, once the model has learned a coarse representation of the scene.
During the subsequent densification stage, where Gaussians may be split or cloned, newly created Gaussians inherit the group identity of their origin Gaussian to preserve grouping consistency. We treat the following as hyperparameters: (i) the number of neighbors for computing each Gaussian’s geometric descriptors (\( l_i \), \( s_i \), \(v_i \), \( p_i \), etc.), and (ii) the grouping strength threshold in Cut Pursuit. These can be adjusted per dataset or reconstruction setting. We found that the effective number of SuperGaussian groups is about 0.02\(\%\)–0.08\(\%\) of the initial number of Gaussians in a scene (for these datasets), corresponding to approximately 16–64 groups. We use $\lambda_1=\lambda_2=1$ for the two regularization terms, and set the SSIM blending factor $\lambda=0.8$ (following prior work). We set $\lambda_{pos}=0.2$. A mask loss is used for Blender/DTU (background masking), with $\lambda_{mask}=0.1$. 

\paragraph{Comparing methods}
We compare our method against 3DGS \cite{kerbl20233d}, its sparse-view variant \cite{xu2024mvpgs,li2024dngaussian,mihajlovic2024splatfields}, and SOTA NeRF-based methods designed for sparse input settings \cite{yang2023freenerf,jain2021putting}. For evaluation, we adopt the metrics reported in the original papers when available; otherwise, we use the official implementations to compute the performance.

\subsection{Comparison}

\paragraph{Blender}
Qualitative and quantitative results on the Blender dataset using 3 and 6 input views are presented in Table~\ref{tab:blender_quant} and Figure~\ref{fig:blender_visual}. COSMOS achieves the best performance across all metrics in the highly sparse 3-view setting, outperforming recent SOTA methods. In particular, the depth map renderings demonstrate that COSMOS constructs continuous geometry and effectively suppresses floaters and noise without relying on any external depth priors. Moreover, even with only 6 input views, COSMOS achieves competitive performance with recent SOTA methods, confirming that the supergaussian‑based 3D priors are effective even under sparse inputs.

\paragraph{DTU}
Qualitative results and visualizations using 3 input views on the dataset are presented in Table~\ref{tab:dtu_quant} and Figure~\ref{fig:dtu_qual}. COSMOS achieves the best performance in terms of PSNR and shows competitive results in SSIM compared to recent methods. Although DNGaussian attains the best SSIM and LPIPS values, our visual results show that COSMOS reconstructs more complex, high-frequency structures without introducing floaters. These results indicate that learning with 3D priors via Gaussian grouping effectively preserves fine-grained geometric details.

Following prior work \cite{somraj2023vip,somraj2023simplenerf}, we evaluate absolute depth prediction performance using the mean absolute error (MAE) of per‑pixel depth. In addition, we assess relative depth prediction performance using the Spearman Rank‑Order Correlation Coefficient (SROCC) \cite{corder2014nonparametric}, which measures the correlation between predicted and ground‑truth pixel depths. The predicted depths of 3DGS trained with dense input views are used as pseudo‑ground truth. As shown in Table~\ref{tab:depth_eval}, COSMOS achieves the lowest MAE and competitive SROCC . These results indicate that, despite not relying on depth supervision, COSMOS effectively predicts depth compared to depth‑based methods such as DNGaussian and MVPGS, demonstrating the benefit of incorporating supergaussian‑based 3D priors.

\begin{table}[t]
  \centering
  \caption{\textbf{Evaluation of depth maps from two depth‑based comparison models on the DTU dataset with three input views.} 
  The reference depth is obtained using 3DGS trained with dense input views.}
  \label{tab:depth_eval}
  \begin{tabular}{lcc}
    \toprule
    Method & MAE ↓ & SROCC ↑ \\
    \midrule
    DNGaussian     & 0.53 & 0.12 \\
    MVPGS          & 0.28 & \textbf{0.44} \\
    COSMOS (Ours)  & \textbf{0.22} & 0.41 \\
    \bottomrule
  \end{tabular}
\end{table}

\subsection{Ablation Study}

\begin{figure}[!htbp]
  \centering
    \centering
    \includegraphics[width=\linewidth]{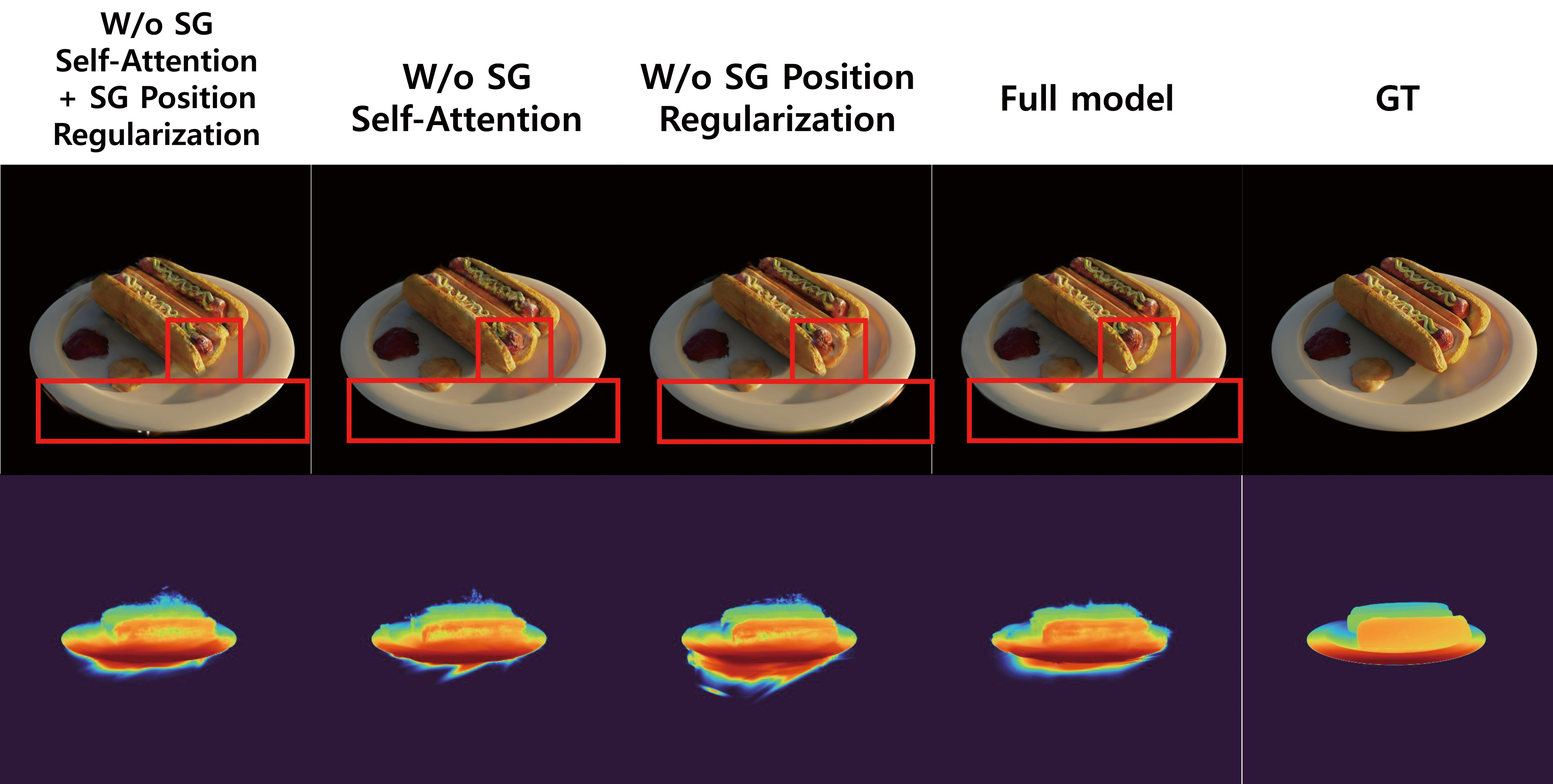}
    \caption{\textbf{ Qualitative Ablation Study of COSMOS. }The red boxes highlight that the Full model captures the most detailed textures. In the depth map renderings, the object surfaces show that the Full model maintains the most consistent and continuous geometry.}
    \label{fig:ablation_visual}
\end{figure}

In this section, we perform an ablation study on the 3‑view Blender dataset to analyze the effectiveness of COSMOS.
As shown in Table~\ref{tab:ablation_ours} and Figure~\ref{fig:ablation_visual}, removing the supergaussian self-attention module leads to significant performance degradation, demonstrating that learning global inter-group features is crucial for producing structurally coherent Gaussian predictions.
Similarly, removing the supergaussian position regularization results in the emergence of noise and floaters around object boundaries, which our method effectively suppresses.
Overall, these ablations confirm that both the supergaussian grouping/self-attention and the positional regularization are critical for robust structure reconstruction under sparse views.

\begin{table}[t]
  \centering
  \caption{\textbf{Quantitative Ablation Study of COSMOS.} SG denotes supergaussian.}
  \label{tab:ablation_ours}
  \begin{tabular}{lccc}
    \toprule
    Method & PSNR ↑ & SSIM ↑ & LPIPS ↓ \\
    \midrule
    w/o SG Self-Attention   & 18.74 & 0.77 & 0.20 \\
    w/o SG Regularization   & 18.24 & 0.76 & 0.21 \\
    w/o SG Regul + SG SA    & 18.22 & 0.76 & 0.21 \\
    Full model              & \textbf{18.96} & \textbf{0.79} & \textbf{0.19} \\
    \bottomrule
  \end{tabular}
\end{table}

\subsection{Limitation}
Setting the optimal number of supergaussians may require cross‑validation. An improper choice could under‑segment or over‑segment the scene. In future work, an adaptive grouping strategy could alleviate this sensitivity to the grouping parameter.

\section{Conclusion}
In this work, we proposed COSMOS, a novel framework that improves 3DGS under sparse-view conditions by grouping Gaussians into supergaussians based on local geometric and appearance cues. We design a representation learning mechanism that integrates both global and local spatial information, enabling structure‑aware prediction of Gaussian attributes (position, rotation, etc.). Furthermore, intra‑group positional regularization is applied to suppress floaters and maintain structural coherence.
 Extensive experiments on the Blender and DTU benchmarks demonstrate that COSMOS consistently outperforms existing methods in few-shot 3D reconstruction.

The idea of supergaussian-based grouping opens up avenues beyond novel view synthesis — for instance, interactive 3D editing or semantic segmentation of Gaussian splats could benefit from our structured priors. We hope this work encourages further exploration of structured priors in point-based 3D reconstruction.

\section{Acknowledgments}
This work was supported by Institute of Information \& communications Technology Planning \& Evaluation (IITP) grant funded by the Korea government (MSIT) (No.RS-2024-00459618, Research on improving intelligent command and control capabilities based on generative AI and real-time 3D digital twin construction) and Next-Generation Intelligent Patrol Platform (Mobile police station) Program through the Korea Institutes of Police Technology (KIPoT) funded by the Korean National Police Agency. (No. RS-2025-25393280)

\bigskip
\noindent 

\bibliography{aaai2026}

\end{document}